\documentclass[12pt]{article}
\usepackage{titlesec}

\titleformat{\section}{\normalfont\Large\bfseries}{\thesection}{0.3em}{}
\titleformat{\subsection}{\normalfont\large\bfseries}{\thesubsection}{0.3em}{}
\titleformat{\subsubsection}{\normalfont\normalsize\bfseries}{\thesubsubsection}{0.3em}{}
\usepackage[nosort,noadjust]{cite}
\usepackage{amsmath,amsthm,amsfonts,amssymb,amscd,mathrsfs,slashed,graphicx,braket}
\usepackage{hyperref}
\usepackage[dvipsnames]{xcolor}
\usepackage{bm}
\hypersetup{
	colorlinks=true,
	citecolor=blue,
	linkcolor=blue,
	urlcolor=blue
}

\newlength{\xtrawidth}
\newlength{\xtraheight}
\numberwithin{equation}{section}
\numberwithin{table}{section}
\numberwithin{figure}{section}
\def\IC{\mathbb{C}}
\def\IP{\mathbb{P}}
\def\IZ{\mathbb{Z}}
\def\IQ{\mathbb{Q}}

\def\DL{\mathfrak{L}}

\def\q{{\mathfrak q}}

\def\qcor#1{\left\langle #1 \right\rangle}

\def\cx#1{{\cal#1}}
\def\tx#1{{\tilde #1}}

\def\tr{\textrm{tr}\,}
\def\Xcf{X^{\sharp}}
\def\ellf{\mathbf{\ell}}
\def\CS{\textrm{CS}}
\def\Li{\textrm{Li}}
\def\Liq{{\textrm{\tt Li}}^q}
\def\tLiq{\widetilde{\textrm{\tt Li}}^q}
\def\LL{\textrm{L}_2}

\def\Pih{\Pi^{\rm \cx S}}
\def\IIb{\II_{\textrm B}}
\def\gs{\lambda}
\def\II{\mathbb{I}}
\def\Fgv{\cx F_{\rm GV}}

\begin{document}
\pagenumbering{Alph}
\begin{titlepage}
\vspace*{-2.5cm}
\begin{center}
\hfill MITP/24-048
\vskip 0.6in
{\LARGE\bf{\strut Non-Perturbative Corrections to}}\\[3ex]
{\LARGE\bf{\strut 3d BPS Indices and Topological Strings}}
\vskip 0.5in
{\bf
Hans Jockers}
\vskip 0.2in
{PRISMA+ Cluster of Excellence \& Mainz Institute for Theoretical Physics\\
Johannes Guttenberg-Universit\"at\\
Staudinger Weg 7, 55128 Mainz, Germany}
\vskip 0.2in
{\tt jockers@uni-mainz.de}
\end{center}
\vskip 0.5in
\begin{center} {\bf Abstract} \end{center}

For a 3d gauged linear sigma model parametrized by a K\"ahler manifold $X$, the 3d BPS index defines a $q$-series that can be analytically continued in the K\"ahler modulus by standard methods. It is argued that an $SL(2,\IZ)$-transform of the Birkhoff connection matrix captures non-perturbative corrections to the 3d GLSM. As an application, a 3d lift of the standard 2d GLSM for the resolved conifold is shown to provide a world-volume dual for the non-perturbative topological string on the resolved conifold. The perturbative 3d BPS index computes the Gopakumar--Vafa partition function, while the analytic continuation matches existing proposals for a non-perturbative completion of the topological string.

\vfill
\noindent May 2024
\end{titlepage}

\pagenumbering{gobble}
\tableofcontents
\newpage
\pagenumbering{arabic}

\section{Introduction and Summary}
To a 3d $\cx N=2$ supersymmetric gauge theory one may assign various BPS indices depending on a fugacity $q$ measuring a combination of spin and $R$-charge, and a number of additional fugacities $y_a$ for the global symmetries. The latter contain a real mass terms for the fundamental fields, such that for generic values of $y_a$ the theory has isolated massive vacua. The starting point of this note is the index of refs.~\cite{BDP,Gadde:2013wq,Gadde:2013sca,DGP,Costello:2020ndc,Dedushenko:2023qjq}, which is equivalent to a 3d partition function on $S^1\times D^2$ under certain conditions \cite{YS,Benini:2015noa,DGP,Bullimore:2020jdq}. The $S^1$ compactified supersymmetric theories of this type have played an important role in a related context of the Bethe/gauge theory correspondence in refs.~\cite{NekSha,Nekrasov:2009uh} and have been connected to quantum K-theory in ref.~\cite{NekPhd}.

In a non-equivariant limit of the 3d index one obtains a class of theories with Higgs branches parametrized by a K\"ahler manifold $X$, which is possibly a Calabi--Yau manifold. The 3d theory compactified on $S^1$ is a certain 3d lift of the well-known 2d gauged linear sigma model (GLSM) for $X$ of refs.~\cite{WitPhases,MP}. In this way the massless limit of the 3d index assigns a set of generalized hypergeometric $q$-series $\Pi_X(Q,q)$ to a K\"ahler manifold $X$, which depends on a choice for the 3d Chern--Simons terms \cite{Jockers:2018sfl}. The fugacities $Q$ represent the complexified K\"ahler moduli of $X$ and are associated to the topological global symmetries in the 3d gauge theory.

In the limit of small radius for $S^1$, the $q$-series $\Pi_X(Q,q)$ reduce to what is usually called the periods of the mirror manifold $Y$ of $X$, if $X$ is Calabi--Yau manifold. In this sense the $q$-series $\Pi_X(Q,q)$ are certain $q$-dependent deformations of the periods of (the mirror of) $X$. Moreover these ``$q$-periods'' $\Pi_X(Q,q)$ satisfy a set of difference equations, which are 3d lifts of the differential GKZ systems well-known from 2d mirror symmetry \cite{Givental:2015zm,Jockers:2018sfl}. 

As for differential equations, the formal $q$-series $\Pi_X(Q,q)$ defined as solutions to the GKZ type $q$-difference equation can be analytically continued in the K\"ahler moduli $Q$, but with some notable differences compared to the 2d case:  the entries of the so-called Birkhoff connection matrix $P_X(Q,q)$ for the analytic continuation are not constants, but elliptic functions on a torus with complex structure $\tau = \ln q/2\pi i$. In this note it is argued, that after an $S$-transformation, the matrix $P_X(Q,q)$ of elliptic functions associated to $X$ captures non-perturbative corrections to the 3d GLSM for $X$.

As a check on these ideas we consider a 3d lift of the standard 2d GLSM for a simple Calabi--Yau manifold, the so-called resolved conifold. The latter is a 2d $\mathcal{N}=(2,2)$ supersymmetric $U(1)$ gauge theory with four chiral multiplets $\Phi_i$, $i=1,\ldots,4$, of charge
\begin{equation}\label{matter}
  \q=(1,1,-1-1)\ .
\end{equation}
This theory has a geometric Higgs branch parametrized by the local Calabi--Yau 3-fold $\Xcf=O(-1)^{\oplus 2}_{\IP_1}$. It describes the local geometry of a generic $\IP^1$ in a general Calabi--Yau 3-fold $X$. The sum over world-sheet instantons computes the Gromov--Witten invariants of $X$ \cite{Witten:1988xj}.  An all genus resummation of these invariants has been obtained in refs.~\cite{Gopakumar:1998ii,Gopakumar:1998jq} from the duality to M-theory compactified on $\Xcf\times S^1$ using a Schwinger type target space computation for 5d BPS states. The result is the Gopakumar--Vafa (GV) potential, which for the conifold takes the form
\begin{equation}\label{FGV}
  \cx \Fgv(t)=\sum_{n=1}^\infty \frac{Q^n}{n(2\sin\frac\gs2)^2}=\sum_{n=1}^\infty \frac{(qQ)^n}{n(1-q^n)^2}\ ,
  \qquad q = e^{i\gs}\ .
\end{equation}
Here $\gs$ is the string coupling constant and $Q=e^{2\pi i t}$ the single complexified K\"ahler modulus of $\Xcf$. Our first observation is, that the above mentioned 3d BPS index for the $S^1$ compactified 3d $U(1)$ GLSM with the same matter content \eqref{matter}, and a particular choice for the 3d Chern--Simons levels, reproduces the first derivative $\partial_t\Fgv(t)$ of the all genus result . It is worth emphasizing that from the point of the 3d gauge theory, $\Fgv(t)$ represents a counting function for BPS operators on the 3d world-volume, in contrast to the 5d target space interpretation of the GV invariants in the sum \eqref{FGV}. In addition $\partial_t\Fgv(t)$ is identified by the 3d GLSM with a generating function of quantum K-theoretic correlators, i.e., virtual holomorphic Euler numbers on the moduli space of stable maps.

Turning to the non-perturbative sector, we find that the Birkhoff connection matrix for this 3d GLSM reproduces the proposal for the non-perturbative completion of the topological string partition function $\Fgv$ put forward in refs.~\cite{Lockhart:2012vp,Hatsuda:2013oxa,Grassi:2014zfa,Alim:2021ukq}.\footnote{For an relation of these proposals to Borel summation, see refs.~\cite{Pasquetti:2010bps,Grassi:2022zuk,Alim:2022oll}. The relation between a non-perturbative index of 3d BPS states and the topological string in this note is reminiscent of that in ref.~\cite{Lockhart:2012vp}, but not identical. The 3d GLSM used in this note is a UV model with a different matter spectrum, and it computes the closed string amplitudes.} The fugacity $q$ is identified with the exponential of the string coupling constant, and the 3d BPS states in the perturbative and the $S$-dual channel map individually to world-sheet and membrane instanton contributions of ref.~\cite{Hatsuda:2013oxa}, respectively.

The potential $\Fgv(t)$ is derived from membrane corrections in the M-theory lift of the topological string. The $S^1$ compactified 3d membrane world-volume gives the fundamental string, which is described by the IR limit of the RG flow of the 2d GLSM \cite{WitPhases}, which in turn arises as a small radius limit of the 3d GLSM we consider. It would be interesting to close the circle of ideas by understanding the details of the RG flow directly in the 3d world volume theory.

\section{Perturbative 3d GLSM \label{sec:GLSM}}
\subsection{Bilateral $q$-series from 3d BPS index}
The $\mathcal{N}=(2,2)$ supersymmetric 2d GLSM provides a weakly coupled gauge theory model for the non-linear sigma model on a K\"ahler manifold $X$ \cite{WitPhases,MP}. The geometry $X$ arises as a Higgs branch of the theory at generic large K\"ahler moduli. In this note we consider a 3d lift of the 2d GLSM for $X$ with the same matter content. If the theory is compactified on $S^1$, it can be described as a 2d gauge theory with infinitely many fields, see, e.g., refs.~\cite{Nekrasov:2009uh, AHKT}.

To the 3d GLSM one can associate several generalized $q$-hypergeometric series~$I(q)$, which describe a certain index. In this note we start from the 3d BPS index of boundary operators defined in refs.~\cite{BDP,Gadde:2013wq,Gadde:2013sca,DGP,Costello:2020ndc}
\begin{equation} \label{DefIIb}
\IIb(y,q)=\tr_{\textrm B}(-1)^{F} q^{J+\tfrac R 2} y^f\,.
\end{equation}
Here $y$ and $f$ stand collectively for the fugacities of the global symmetries and the associated charges, and the parameter $q$ is the fugacity for the indicated combination of the spin $J$ and $U(1)_R$ symmetry quantum numbers. The index depends on a choice of boundary conditions on the fields, and the Chern--Simons levels for the local and global symmetries, as indicated by the subscript  B.

To keep the discussion simple, we restrict to the case of a $U(1)$ gauge theory with chiral matter fields of $U(1)$ charges $\q_\alpha$, although more general gauge groups and matter representations can be treated similarly. Taking Dirichlet boundary conditions for the gauge fields, the 3d index $\IIb$ takes the schematic form \cite{DGP}
\begin{equation}\label{3dindex}
\begin{aligned}
  \IIb(Q,s,q)&=\frac{1}{(q;q)_\infty}\sum_{m\in\IZ}Q^{-m} \Upsilon(m)\ ,\\
  \Upsilon(m)&= q^{\frac {A}2 m^2+Bm}s^{A m}\times[\textrm{matter index}](q^ms)\ .
\end{aligned}
\end{equation}
Here $(Q,s)$ are the fugacities of the topological $U(1)_t$ symmetry dual to the $U(1)$ gauge group and the global symmetry $U(1)_\partial$ on the boundary, respectively. The sum over $m$ takes into account the sectors with different monopole number. Moreover, $A$ and $B$ denote the effective Chern--Simons levels for the gauge/gauge and gauge/R-symmetry Chern--Simons action.\footnote{The Chern--Simons levels include one-loop corrections and boundary terms \cite{DGP,YS}, as opposed to the Chern--Simons levels in the classical action.} The indices of the matter fields of $U(1)$ charge $\q$, $U(1)_R$ charge $r$ and N(eumann)/D(irichlet) boundary conditions are \cite{Gadde:2013wq,Gadde:2013sca,DGP}
\begin{equation}\label{chiralindex}
N: \ (\lambda s^\q q^{r/2};q)^{-1}_\infty\,,\qquad 
D: \ (\lambda s^{-\q}q^{1-r/2};q)_\infty\,,
\end{equation}
with the $q$-Pochhammer symbol defined as $(z;q)_\infty=\prod_{r=0}^\infty (1-zq^r)$ for $|q|<1$. The argument $\lambda$ denotes an additional fugacity for a global symmetry rotating the chiral multiplets. For the 3d lift of a 2d GLSM for a complete intersection of degree $d_i$ hypersurfaces in a toric variety $W$ we consider the spectrum in Table~\ref{tab:spec}.

\begin{table}[t]
\begin{center}
\begin{tabular}{| l | c | c | l |}
\hline
& $U(1)$ charge & $U(1)_R$ charge & Boundary cond.\\ 
\hline
Toric variety $W$ & $\phantom{+}\q_\alpha$, \ $\alpha=1,\ldots,n$ & $r=0$ & \ Neumann b.c. \\
Hypersurface constraints & $-d_i$, \ $i=1,\ldots,m$ & $r=2$ & \ Dirichlet b.c. \\
\hline
\end{tabular}
\end{center}
\caption{The spectrum for a 3d GLSM with a Higgs branch parametrizing complete intersection of degree $d_i$ hypersurfaces in a $n$-dimensional toric variety $W=\IC^n/\!\!/\IC^*$ of co-dimension $m$.} \label{tab:spec}
\end{table}

In addition one has to specify the values of the Chern--Simons levels in the lifted theory. After some rearrangements, the 3d index for this spectrum takes the form of a bilateral $q$-hypergeometric series 
\begin{equation}
\IIb(Q,s,q)=N\cdot (\II_++\II_-)\,,
\end{equation}
where $ N=(q;q)^{-1}\prod_i(s^{d_i};q)_\infty\prod_\alpha(s^{\q_\alpha};q)^{-1}_\infty$ and 
\begin{equation} \label{IIser}
\begin{aligned}
 \II_- &= \sum_{m\geq 0} \tilde Q^{m}s^{-m\tilde A} q^{\frac m2 \tilde A^2 -\tilde Bm}\, \frac{\prod_{\q_\alpha<0}(s^{-\q_\alpha}q^{1+m\q_\alpha};q)_{m|\q\alpha|}}{\prod_{\q_\alpha>0}(qs^{-\q_\alpha};q)_{m\q_\alpha}}\prod_{i}(s^{-d_i}q;q)_{md_i}\ , \\
\II_+ &= \sum_{m>0} \tilde Q^{-m} s^{m\tilde A} q^{\frac m2 \tilde A^2 +\tilde Bm}\,\frac{\prod_{\q_\alpha>0}(s^{-\q_\alpha}q^{1-m\q_\alpha};q)_{m\q\alpha}}{\prod_{\q_\alpha<0}(s^{-\q_\alpha}q;q)_{m|\q_\alpha|}}\frac 1{\prod_{i}(s^{-d_i}q^{1-md_i};q)_{md_i}}\ .
\end{aligned}
\end{equation}
Here we used the redefinitions $\tx Q = (-1)^{c_1}Q$, $\tx B=B+c_1/2$ and $\tx A = A+\textrm{ch}_2$, where $c_1=\sum_\alpha \q_\alpha-\sum_i d_i$ and $\textrm{ch}_2= \sum_\alpha\q_\alpha^2-\sum_id_i^2$ are the numerical coefficients of the first and second Chern characters of $X$. The tildes are dropped again below.

A concise description of this $q$-series is as a solution to a difference equation.\footnote{In the 3d gauge theory these equations represent Ward identities of line operators \cite{Dimofte:2011py,KW13}.}  Generally, for a 3d theory with a Higgs phase parametrized by a toric manifold $X$, there is a canonical set $\{\DL_a\}$ of difference operators \cite{Jockers:2018sfl}, which constitute a 3d lift of the well-known  GKZ differential system $\{L_a\}$ for the 2d GLSM for $X$. The latter is relevant for quantum cohomology and annihilates the period integrals of the mirror of $X$.\footnote{For zero Chern--Simons level, these difference equations have first appeared in ref.~\cite{Givental:2015ef} in the context of quantum K-theory. For background material and references on quantum cohomology, see, e.g., ref.~\cite{CK}.} The basic difference operator $p$ for the $U(1)$ gauge theory shifts the topological fugacity $Q$ as
\begin{equation}\label{defp}
 p\, Q=qQ\, p\,.
\end{equation}
Then the bilateral series  $\II=e^{-\frac{\ln Q \ln s}{\ln q}}(\II_-+\II_+)$ satisfies the difference equation $\DL \II =0$, where $\DL$ is of the generalized $q$-hypergeometric form \cite{Jockers:2018sfl,Jockers:2021omw}
\begin{equation}\label{DL}
\DL =  
\prod_{\q_\alpha>0}\prod_{j=0}^{\q_\alpha-1}(1-p^{\q_\alpha}q^{-j})-Qq^{\frac A2 -B}p^{A} 
\prod_{\q_\alpha<0}\prod_{j=0}^{|\q_\alpha|-1}(1-p^{\q_\alpha}q^{-j})\, \prod_{i}\prod_{j=1}^{d_i}(1-p^{d_i}q^{j}) \, .
\end{equation}
The prefactor $e^{-\frac{\ln Q \ln s}{\ln q}}$ simplifies the difference equations, by avoiding extra factors of $s$. It represents a mixed Chern--Simons term for the gauge/topological $U(1)$ symmetries, which changes the analytic properties of the solution, as it is not periodic under $s\to e^{2\pi i} s$ and $Q\to e^{2\pi i }Q$. One can replace it by another function $f(Q)$ with the same transformation behavior $pf(Q) = sf(Q)$, in particular a product of theta functions, as discussed in Section~4.2 of ref.~\cite{BDP}. We come back to this point in the next section.

As it stands, $\IIb$ is a formal series in the fugacities $(Q,s,q)$ with no obvious convergence properties. In particular it is a bilateral series with positive and negative powers of $Q$. To obtain a geometric interpretation related to a Higgs vacuum with target $X$, one has to identify the $U(1)_\partial$ fugacity $s$ with the K-theoretic Chern class $P^{-1}=e^{c_1(O(1))}$ of the hyperplane bundle \cite{Jockers:2021omw,Jockers:2018sfl}. Expanding the resulting K-theory valued series in $H=1-P$ one gets
\begin{eqnarray}
\IIb(Q,s=P^{-1},q)=N I_X(Q,q)+ O(H^{\dim(X)+1})\,,
\end{eqnarray}
where $N$ is a suitable normalization factor used to normalize the power series $I_X$ in $Q$. Imposing the K-theory relation $H^{\dim(X)+1}=0$, the bilateral series $\IIb$ collapses to the K-theory valued power series $I_X=\II_-|_{H^{\dim(X)+1}=0}$. The series $I_X$ is a solution of the same difference equation, $\DL I_X=0$, up to terms of $O(H^{\dim X+1})$ and will serve as a starting point for the analytic continuation.\footnote{The series~$I_X$ can be viewed as a vortex sum of the type studied in ref.~\cite{DGH}. The reduction to the power series reflects the restriction of the index to operators compatible with the boundary conditions imposed by the vacuum, as discussed in ref.~\cite{Verma} in a similar context.}

The $S^1$ compactified 3d gauge theory has a K-theoretic interpretation \cite{NekPhd,NekSha,Nekrasov:2009uh,KW13}. The quantum K-theory relevant for the $\cx N=2$ supersymmetric 3d GLSM has been identified in ref.~\cite{Jockers:2018sfl} as  the permutation equivariant theory defined in ref.~\cite{Givental:2015bf}, and its higher level generalization \cite{RZ18}. In this context, $I_X$ is a generating function of K-theoretic correlators defined by holomorphic Euler numbers on the moduli space of stable maps, possibly with a non-trivial (permutation equivariant) input \cite{Jockers:2018sfl,Garoufalidis:2021cha}.\footnote{See the earlier and related works \cite{NekPhd,Bullimore:2014awa,Pushkar:2016qvw,Koroteev:2017nab,Dedushenko:2023qjq} for a UV definition of the correlators  corresponding to quasimap moduli spaces.} Using reconstruction methods \cite{Givental:2015VIII}, such an input can be removed and the holomorphic Euler numbers of the resulting correlators generate the K-theoretic Givental J-function. The necessary transformation to remove a non-trivial input is often referred to as the mirror map of quantum K-theory.

\subsection{A 3d GLSM dual for the topological string on the conifold}
We now apply the above ideas to the 3d GLSM with the spectrum \eqref{matter}. In this case there is no hypersurface constraint. Taking Neumann boundary conditions in all four directions the index is  
\begin{equation}
\II = \frac1{(q;q)_\infty}\sum_{m\in\IZ} \frac{q^{\frac 12 A m^2+Bm}s^{Am}Q^m}{(sq^m\lambda_+;q)^2_\infty(s^{-1}q^{-m}\lambda_-;q)^2_\infty}\,.
\end{equation}
Here we have introduced temporarily non-trivial fugacities $\lambda_\pm$ for the two pairs of matter fields of the same charge to regularize the non-compact directions. This amounts to giving a real mass to the 3d chirals and computing equivariantly with respect to the $(\IC^*)^2$ symmetry that acts by phase rotations on the two sets of fields.

For the conifold the two parts of the bilateral sum with $m<0$ or $m>0$ represent the vortex sums of two different Higgs vacua centered around $Q=0$ and $w = Q^{-1}=0$, respectively, e.g., the negative part can be written as
\begin{equation} \label{Imcf}
\begin{aligned}
\II_-&=N\sum_{m>0} Q^m q^{\frac 12 A m^2-Bm}s^{-Am}\frac{\prod_{r=0}^{m-1}(1-s^{-1}\lambda_-q^r)^2}{\prod_{r=1}^{m}(1-s\lambda_+q^{-r})^2}\\
&=
N(1-s\lambda^{-1}_-)^2 \sum_{m>0} Q^m q^{\frac 12m^2(A+2)+m(1-B)}s^{-(A+2)m-2}\lambda_+^{-2m}\lambda_-^2\frac{\prod_{r=1}^{m-1}(1-s^{-1}\lambda_-q^{r})^2}{\prod_{r=1}^{m}(1-s^{-1}\lambda^{-1}_+q^{r})^2}\ ,
\end{aligned}
\end{equation}
with $N^{-1}=(q;q)_\infty(s^{-1}\lambda_-;q)^2_\infty(s\lambda_+;q)^2_\infty$.

The Higgs phase near $Q=0$ exists in the massless limit $\lambda_+\to 1$. Identifying the boundary fugacity $s$ with the inverse of the Chern character $P$ of the line bundle $O(-1)_{\IP^1}$, the prefactor $(1-s\lambda^{-1}_-)^2$ in $\II_-$ becomes the equivariant K-theoretic Euler class $e_{S^1}^K$ of the normal bundle $O(-1)\oplus O(-1)$ to $\IP^1$. The positive sum $\II_+$ is obtained from \eqref{Imcf} by the replacements $(Q,s,\lambda_\pm,B)\to(Q^{-1},s^{-1},\lambda_\mp,-B)$. The prefactor $(1-P\lambda_+)^2$ in $\II_+$ is zero in the limit $\lambda_+=1$ by the K-theory relation $(1-P)^2=0$ for $\IP^1$. Stripping off  the normalization factor $N$ and taking the non-equivariant limit $\lambda_\pm\to 1$ one obtains for the index in the Higgs vacuum near $Q=0$
\begin{equation}\label{IK}
I_X(Q,q)=1+e_{S^1}^K\left(\sum_{m>0}\frac{Q^dP^{2+m(A+2)}q^{\frac {m^2}2(A+2)+m(1-B)}}{(1-Pq^m)^2}+O(1-\lambda_-)\right) \,.
\end{equation}
For Chern--Simons levels $(A,B)=(-4,0)$ this agrees with the K-theoretic $I$-function for the resolved conifold of ref.~\cite{Givental:2015zm}. \\

\subsubsection{Difference equation and local solutions}
To write the difference operator for $I_X$ on a non-compact Calabi--Yau manifold such as the resolved conifold, one needs to regularize the solutions associated with the non-compact elements of $H^{2*}(X)$. This can be done either by a mass deformation of the non-compact directions, or by embedding of the local Calabi--Yau manifold into a compact Calabi--Yau manifold to compactify the relevant cycles.\footnote{For similar constructions in the context quantum cohomology and mirror symmetry of non-compact Calabi--Yau threefolds, see for instance the discussions in refs.~\cite{Chiang:1999tz,Forbes:2005xt}.} For the resolved conifold~$\Xcf$ with standard Chern--Simons levels this has been already discussed in refs.~\cite{Givental:2015zm, Jockers:2019wjh}. Since the 3d theory dual to the topological string has different Chern--Simons levels, we first need the difference equation for general levels. 
  
It is again useful to multiply the vortex sum by a prefactor representing a mixed Chern--Simons term
\begin{equation}\label{pref}
P^{\ellf_Q}=\sum_{k=0}{\ellf_Q\choose k}(-H)^k\,, \qquad H=1-P\,,
\end{equation}
where $\ellf_Q$ satisfies the difference equation $p\,\ellf_Q=\ellf_Q+1$. One may take $\ellf_Q=\frac{\ln (\alpha Q)}{\ln q}$,  as is used with the choice $\alpha=1$ before eq.~\eqref{DL}, but there are other choices differing by shift invariant functions. The prefactor defines an invertible transformation in $K(\Xcf)$ and avoids a $P$-dependence of the difference operator due to the relation $(1-p)(P^{\ellf_Q}I)=P^{\ellf_Q}(1-Pp)I$. Relabeling the Chern--Simons levels as $(a,b)=(A+2,-B+1)$, we are led to consider the series
\begin{equation}\label{DefI}
I=P^{\ellf_Q}(1-P)^2\sum_{m\geq 0}\frac{Q^mq^{\CS(m)}P^{am}}{(1-Pq^m)^2}\ ,\qquad q^{\CS(m)}=q^{\frac a2 m^2+bm} \ .
\end{equation}
The $q$-deformed GKZ operator $\DL$ that annihilates $I$ is 
\begin{equation}\label{diffop}
  \DL=(1-p)^2(1-Qq^{\frac a2 +b}p^a)(1-p)^2\ .
\end{equation}
This operator reduces to that of ref.~\cite{Jockers:2019wjh} for Chern--Simons levels $(a,b)=(-2,1)$.

The prefactor in eq.~\eqref{pref} implements the Frobenius method for the solutions to the difference equation $\DL f(Q)=0$ near $Q=0$. These are the first four terms in an expansion
\begin{equation} \label{IQexp}
I(Q,q)=\sum_{a=0}^3 I_a(Q,q) H^a + O(H^4)\ .
\end{equation}
The first coefficients multiply the four generators of the topological K-theory of $\Xcf$, or by the Chern isomorphism, $\oplus_{a=0}^3 H^{2a}(\Xcf,\IQ)$. The terms proportional to $H^2$ and $H^3$ correspond, in cohomology, to elements dual to the non-compact 4- and 6-cycles in the resolved conifold~$\Xcf$. The vanishing of the higher order terms is due to the ideal $(1-P)^2(1-\lambda_-P)^2=0$ of the toric space in the non-equivariant limit $\lambda_-\to 1$.\footnote{A more detailed discussion of the regularization of the non-compact directions can be found in ref.~\cite{Jockers:2019wjh}.}

The expansion of the sum over $Q$ in eq.~\eqref{DefI} in $H$ is $1+\Liq_2(Q)H^2-2\Liq_3(Q)H^3$ with\footnote{The notation indicates that these functions are $q$-deformations of the polylogarithms.}
\begin{equation}\label{defi2}
\Liq_2(Q) = \sum_{n>0}Q^nq^{\CS(n)}\frac{1}{(1-q^n)^2}\, ,\qquad 
\Liq_3(Q) = \sum_{n>0}Q^nq^{\CS(n)}\, \frac{\frac a2 n(1-q^n)+q^n}{(1-q^n)^3}\,.
\end{equation}
The four-dimensional solution vector $\Pi_0(Q,q)$ with components $I_a$ at $Q=0$ written in terms of these functions is  

\begin{equation}\label{sol0}
\begin{aligned}
\Pi_0(Q,q) & =
  \begin{pmatrix}1\\-\ellf_Q\\ \frac 12 \ellf_Q(\ellf_Q-1)+\Liq_2(Q)\\
  -\frac1{3!}\ellf_Q(\ellf_Q-1)(\ellf_Q-2)-\ellf_Q\,  \Liq_2(Q)-2\Liq_3(Q)\end{pmatrix} \ , \\
  \Pi_\infty(w,q)&= \begin{pmatrix}1\\-\ellf_w\\  {\tLiq_2}(w)\\
  -\ellf_w\,   {\tLiq_2}(w)-2 {\tLiq_3}(w)\end{pmatrix} \ .
\end{aligned}  
\end{equation}
Similarly, the solution vector $\Pi_\infty$ in the other geometric Higgs phase near $w=Q^{-1}=0$ is determined by  the transformed difference equation 
\begin{equation} \label{eq:DiffwEq}
(1-q^{-1} p_w)^2(1-wq^{\frac a 2+2 -b}p_w^{a})(1-p_w)^2\  f(w)=0\ ,\qquad 
p_w = p^{-1}\ .
\end{equation}
In the expression for $\Pi_\infty$ on the right hand side of eq.~\eqref{sol0},  ${\tLiq_2}(w)$ and ${\tLiq_3}(w)$ are given by eq.~\eqref{defi2} with the shift $(a,b)\to (a,2-b)$ of Chern--Simons levels.

These four-dimensional vectors of solutions in the vicinity $Q=0$ and $w=Q^{-1}=0$ are the respective fundamental solutions of the difference equations~\eqref{diffop} and \eqref{eq:DiffwEq}. They are constructed from the incidence difference equations at $Q=0$ and $w=0$. Depending on the choice of the Chern--Simons levels these difference operators are actually of higher order at $Q=w^{-1} \ne 0$. Therefore, there are further exceptional solutions that can be constructed with the Adam's method, see for instance~\cite{MR4438728}. The fundamental solutions correspond to the K-theory elements of the resolved conifold $X^\sharp$ and are relevant for the analytic continuation discussed in this work. The remaining solutions --- relating to twisted K-theory elements in the inertia stack of quasi-maps \cite{RZ18} --- are not further analyzed here.

The functions in eq.~\eqref{defi2} are certain $q$-deformations of the classical polylogarithms. Generally, writing $q=e^{-\beta\hbar}$, where $\beta$ denotes the $S^1$ radius and $\hbar$ the $U(1)$ weight corresponding to the combination $J+R/2$ in eq.~\eqref{DefIIb}, the $\beta \to 0$ limit of the difference operator $\DL$ gives the GKZ operator for the 2d GLSM with a Calabi--Yau target $X$, and the coefficients $I_a$ of the expansion  $I_X$ reduce to the periods of the mirror manifold of $X$ up to linear combination.\footnote{If $X$ is not Calabi--Yau space, one has in addition to take into account the renormalization of $Q$, see Section~2.1 of ref.~\cite{Jockers:2018sfl}.}

For the special case of the conifold, the difference operator $\DL$ reduces in the small radius limit to the GKZ differential operator $L=\theta^2(1-Q)\theta^2$ annihilating the ordinary periods of the mirror of the resolved conifold. The leading terms of the functions in $\Pi_0$ in an expansion in small $\beta$ are 
\begin{equation}
\ellf_Q\to -\frac1{\hbar\beta}\, \ln (\alpha Q)\ ,\quad
  \Liq_2(Q)\to \frac1{(\hbar\beta)^{2}}\, \Li_2(Q)\ ,\quad
  \Liq_3(Q)\to \frac1{(\hbar\beta)^{3}}\, \Li_3(Q)\ .
  \end{equation}
As a result, the 2d limit of $\Pi_0$ is the period vector for the mirror of the resolved conifold~$\Xcf$, and the $q$-hypergeometric series $I_a$ can be viewed as a (level dependent) $q$-deformation of the classical periods of the mirror manifold. For this reason we refer to the entries of the solution vector $\Pi_0$ as the $q$-periods of the 3d GLSM.\footnote{More precisely, the solution vector $\Pi_0$ and its 2d limit are a linear combination of the entries of the period vector. The 2d period vector comes with an integral structure defined by the cohomology group $H^3(Y,\IZ)$ of the mirror $Y$. There is a similar integral structure defined by quantum K-theory, which reduces to the 2d integral structure in the 2d limit, see, e.g., ref.~\cite{Jockers:2018sfl}.} The relationship between the 2d limiting periods and their deformed $q$-periods as solutions to differential equations and their deformed $q$-difference equations is known as confluence. The notion of confluence for $q$-difference equations is introduced in ref.~\cite{MR1799737}, which is applied to quantum K-theory in refs.~\cite{MR4438728,MR4506744}. 

\subsubsection{Duality to topological string and quantum K-theory}
The $S^1$ compactified 3d gauge theory has a K-theoretic interpretation on general grounds \cite{NekPhd,NekSha,Nekrasov:2009uh,KW13}. The quantum K-theory relevant for the $\cx N=2$ supersymmetric 3d GLSM has been identified in ref.~\cite{Jockers:2018sfl} as  the permutation equivariant theory defined in ref.~\cite{Givental:2015bf}, and its higher level generalization \cite{RZ18}. In this context, the power series $I_X$ is the generating function of certain holomorphic Euler numbers on the moduli space of stable maps.

If $X$ is a Calabi--Yau 3-fold, the series $I_X$ can be expressed in closed form in terms of the Gopakumar--Vafa invariants, according to a proposal in ref.~\cite{Jockers:2019wjh}, which is proven in ref.~\cite{Chou:2023xnn}. The two different series $I_X(Q,q)$ and $\Fgv(Q,q)$ thus contain equivalent information as local power series, although the physical interpretation of the two expansions is different: the series $I_X$ counts BPS operators in the 3d gauge theory, while $\Fgv$ counts 5d BPS states in the target space theory. In this sense quantum K-theory provides a certain resummation of the topological string, with the string coupling constant $q=e^{i\lambda}$ identified with the fugacity $q$ in the world-volume index \eqref{DefIIb}. The classical terms in the large volume limit also match if one takes into account the integral structure of quantum K-theory defined by an elliptic version of D-brane boundary conditions, compare with Section~9 of ref.~\cite{Jockers:2018sfl}.

For the special case of the conifold, we find here a particular simple relation between the 3d BPS index for the above $U(1)$ GLSM model on the resolved conifold~$\Xcf$ with standard spectrum \eqref{matter} and the topological string. There is a choice of Chern--Simons levels, such that the generating function of the quantum K-theoretic correlators coincides with $\partial_t\Fgv$. For the choice of levels 
\begin{equation}
  \label{CSts} (a,b)=(0,b) \ ,
\end{equation} one has
\begin{equation}
\frac{\partial}{\partial t} \Fgv(t+\beta\hbar (1-b))=\Liq_2(Q)=\frac1{1-q}\sum_{d>0}Q^d \qcor{\frac H {1-qL}}_{g=0,\beta=d[\IP^1]}\,,
\end{equation}
where the expression on the right hand side is the quantum K-theoretic correlator as defined in ref.~\cite{Givental:2015bf}.\footnote{Here $g$ denotes the genus and $\beta$ the class of a stable map, as usual, and $d$ is the degree of the map to the compact $\IP^1$ in $\Xcf$.} The match of the 3d and 5d BPS indices indicates that there is a world-volume/target space duality at work. It would be interesting to study the details of the mapping between the moduli spaces of the 3d and 5d BPS states and a possible generalization to other Chern--Simons levels.

\section{Non-perturbative Corrections from Analytic Continuation}
The index counts the BPS operators of the 3d theory irrespective of the vacuum, while the local expansions around different values of the K\"ahler moduli describe the perturbative spectrum compatible with Higgs or Coulomb type vacua. The local solutions to the $q$-difference operators \eqref{DL} can be analytically continued in the complexified K\"ahler moduli $Q$ by standard methods similar to the case of differential equations.\footnote{See refs.~\cite{MR3410204,Rahman} for background material and refs.~\cite{MR4438728,MR4506744} for application to quantum K-theory.} Below we find that the connection terms introduced by the analytic continuation capture non-perturbative corrections to the local perturbative expansions. This relation holds generally for the 3d GLSM with target $X$, in particular for Calabi--Yau target spaces. The general idea is demonstrated in the following for the GLSM for the resolved conifold with the Chern--Simons levels \eqref{CSts} dual to the topological string.

\subsection{Birkhoff connection matrix for the perturbative index}
We first briefly review necessary definitions and facts from the analytic continuation of solutions to $q$-difference equations from refs.~\cite{MR1799737,MR3410204,MR2032530}. Consider a $q$-difference equation $\DL\, \Pi=0$ of order $N$ for the operator $\DL=\sum_{i=1}^N a_i(Q,q)p^i$ , where $pQ=qQp$. This can be rewritten in matrix form as 
\begin{equation}
pX = AX \,,
\end{equation}
where the $n$-th row of the $N\times N$ matrix $X(Q,q)$ is $p^{n-1}\Pi$, where $\Pi$ is the vector of solutions. Denote by $X_0,A_0$ the local expansions near $Q=0$ and $X_\infty,A_\infty$ the expansions near $w=Q^{-1}=0$, respectively. One has
\begin{equation}
p_wX_\infty = A_\infty(w)X_\infty \,,\qquad p_w=p^{-1}\,,\quad A_\infty(w)=A^{-1}_0(z/q)\,.
\end{equation}
The analytic continuation between the solutions $\Pi_0(Q)$ and $\Pi_\infty(w)$ can be written as a matrix equation
\begin{equation}\label{ac}
\Pi_\infty = P^{-1T}\Pi_0\,,
\end{equation}
where $P$ is the Birkhoff connection matrix, given in terms of the $q$-period matrices  as 
\begin{equation}
P=X_\infty^{-1}X_0\,. 
\end{equation}
The connection $P$ is shift invariant, i.e., $P(Qq)=P(q)$. If the solutions $\Pi_0$ and $\Pi_\infty$ are chosen to take values in the appropriate field, it is moreover invariant under $Q\to e^{2\pi i} Q$ and meromorphic, i.e., its entries are elliptic functions on the torus with coordinate $u$ and complex structure $\tau$, where
\begin{equation}
u=\frac{\ln Q}{2\pi i}\ ,\qquad \tau=\frac{\ln q }{2\pi i}\ .
\end{equation}
In this way we get a matrix of elliptic functions associated with the 3d GLSM for $X$.

\subsubsection{Connection matrix for the 3d GLSM for $\Xcf$}
Applied to to the GLSM for the conifold $\Xcf$ with matter content \eqref{matter}, the matrices $X_0$, $X_\infty$ and the Birkhoff matrix can be straightforwardly computed from the $q$-periods $\Pi_0$ and $\Pi_\infty$ in eq.~\eqref{sol0}. Restricting to the Chern--Simons levels \eqref{CSts} dual to the topological string, and setting $b=0$ for simplicity, one obtains
\begin{equation}
P=\begin{pmatrix}
1 & 0 & -R & S\\
 0 & -1 & 0 & -R \\
 0 & 0 & -1 & -2 \\
 0 & 0 & 0 & 1 
\end{pmatrix} \ , \qquad
P^{-1}=\begin{pmatrix}
1 & 0 & -R & -2R - S \\
 0 & -1 & 0 & -R \\
 0 & 0 & -1 & -2 \\
 0 & 0 & 0 & 1 
\end{pmatrix}\ ,
\end{equation}
with the non-constant entries
\begin{equation}
\begin{aligned}
  R&=-\Liq_2(Q)-\Liq_2(wq^2)-\frac 12 \ellf_Q(\ellf_Q-1)\ ,\\
  S&=2(\ellf_Q-1)R+\frac 13 \ellf_Q(\ellf_Q-1)(\ellf_Q-2)+2(\Liq_3(wq^3)-\Liq_3(Q)-\Liq_2(Q))\ .
\end{aligned}
\end{equation}
Here the functions $\ellf_Q$ are defined as \cite{MR3410204}
\begin{equation}\label{preftheta}
  \ellf_Q=-d_Q\ln\theta(Q,q)\ , \quad  \theta(Q,q)=(Q,q)_\infty(q/Q,q)_\infty\ , \quad d_x=x\frac d {dx}\ ,
\end{equation}
$\ellf_Q$ has a first order pole at $Q=q^k$, $k\in\IZ$. Resumming the series $\Liq_2$ and $\Liq_3$ defined in eq.~\eqref{defi2} as
\begin{equation}
\Liq_2(Q)=\sum_{k=1}^\infty (k+1)g_k(Q)\ , \quad
\Liq_3(Q)=\sum_{k=1}^\infty\frac{(k+1)(k+2)}2 g_k(qQ)\ ,
\end{equation}
with
\begin{equation}
  g_k(Q)=\frac{Qq^k}{1-q^kQ}\ ,
\end{equation}
shows that these terms contribute further poles of order 1 and 2, respectively. Thus $R$ and $S$ are double periodic meromorphic functions with poles of order $2$ and $3$, which can be expressed in terms of the Weierstrass function and its first derivative, respectively. The expansion of $\cx P(u,\tau)$ near a pole $u\simeq 0$ is
\begin{equation}
\cx P(u,\tau)=\frac{1}{u^2}+\frac{g_2(q) u^2}{20}+\frac{g_3(q) u^4}{28}+\frac{g_2(q)^2u^6}{1200}+O\left(u^8\right)\, .
\end{equation}
Here $g_2(q)=\frac 4 3 \pi^4 E_4(q)$ and $g_3(q) = \frac 8 {27}\pi^6 E_6(q)$, and $E_k(q)$ are the Eisenstein series. Comparing to the poles in the expansions of $R$ and $S$ one obtains the searched for relations
\begin{equation}
  R= \frac 1 {24}-\frac 12 \frac 1 {(2\pi i)^2}\cx P(u,\tau)\,,\qquad S=-\frac 13  \frac 1 {(2\pi i)^3}\partial_u\cx P(u,\tau)\ .
\end{equation}

It is instructive to have a more explicit derivation of the relation between the local series and  the Weierstrass function. For the $H^2$ term in the expansion \eqref{IQexp} --- loosely speaking the 4-cycle $q$-period --- one shows $\Liq_2(Q)+\Liq_2(q^2w)+\LL(Q,q)=0$ with 
\begin{equation}\label{acH2}
\LL(Q,q)=\frac12 \ellf_Q(\ellf_Q-1)+\frac 1 {24}-\frac 12 \frac 1 {(2\pi i)^2}\cx P(u,\tau)\ .
\end{equation}
Using 
\begin{equation}
\partial_u^2\ln \theta(e^{2\pi i u},q) = -P(u)+ (2\pi i)^2\left( \frac 12 q\partial_q \ln(q)_\infty+\frac 1{12} \right)\ ,
\end{equation}
this function can be rewritten as 
\begin{equation}\label{Xtheta}
  \LL(Q,q)= (d_Q+d_q)\ln \theta(Q,q) \ .
\end{equation}
The identity \eqref{acH2} is then verified by splitting the theta function into the two factors of eq.~\eqref{preftheta} depending only on $Q$ or $w$, respectively, with series expansions
\begin{equation}\label{serexp1}
(d_Q+d_q)\ln (Q,q)_\infty=-\Liq_2(Q)\,,\qquad (d_Q+d_q)\ln (qw,q)_\infty =-\Liq_2(q^2w)\,.
\end{equation}

\subsection{Non-perturbative solutions from $S$-duality}
The matrix identity \eqref{ac} from the analytic continuation of the $q$-periods is expressed in terms of the original variables $Q,q$. The elliptic functions in the entries of the connection matrix can be $SL(2,\IZ)$ transformed. To display the non-perturbative content of the connection matrix $P$, we rewrite it in the $S$-dual channel with variables
\begin{equation}
\tx Q = Q^{\frac 1\tau}\,,\qquad \tx q = e^{-\frac{2\pi i}{\tau}}\,.
\end{equation}

Continuing with the conifold as an illustrative example, the $S$-transformation of the Jacobi theta functions implies 
\begin{equation}
\theta(Q,q)=i(\tx q/q)^{\frac 1{12}}\sqrt{Q/\tx Q}\, e^{-\frac{\ln^2 Q}{2\ln q}}\theta(\tx Q,\tx q)\,.
\end{equation}
In the $S$-dual variables $\LL=\LL^{(1)}+\LL^{(0)}$ with
\begin{equation}\label{Xdual}
\begin{aligned}
\LL^{(1)}(\tx Q ,\tx q)&=\left[\frac 1{2\pi i}\partial_\tau \ln \theta(Q^{\frac 1 \tau},\tx q) \right]_{\tx Q \to \tx Qe^{-2\pi i }}\ ,\\
\LL^{(0)}(Q,q)&=\frac{\ln^2(e^{i\pi}Qq^{-\frac 12})}{2(2\pi i \tau)^2}-
\frac{\ln(e^{i\pi}Qq^{-\frac 12})}{2(2\pi i \tau)}-\frac 1 {24\tau^2}+\frac 1{24}\ .
\end{aligned}
\end{equation}
Using again $\theta(\tx Q,\tx q)=(\tx Q,\tx q)_\infty(\tx w\tx q,\tx q)_\infty$ and a series expansion as in eq.~\eqref{serexp1}, the function~$\LL^{(1)}(\tx Q,\tx q)$ can be split in two parts which can be assigned to expansions near the two different Higgs vacua near $Q=0$ and $w=0$, respectively
\begin{equation}\label{Xplus}
\begin{aligned}
\LL^+(\tx Q,\tx q)&=\left[\frac 1{2\pi i}\partial_\tau \ln (Q^{\frac 1 \tau},\tx q)_\infty \right]_{\tx Q \to \tx Qe^{-2\pi i }}\\
  &=-\frac1{\tau^2} \sum_{n>0}\frac{\tx Q^n\tx q^n}{(1-\tx q^n)^2}-\frac{\ln (\tx Q)-2\pi i}{2\pi i\tau}\sum_{n>0}\frac{\tx Q^n}{1-\tx q^n}\ , \\
\LL^-(\tx w,\tx q)&=\left[\frac 1{2\pi i}\partial_\tau \ln (w^{\frac 1 \tau}\tx q,\tx q)_\infty \right]_{\tx w \to \tx we^{2\pi i }}\ .
\end{aligned}
\end{equation}
These expressions are similar --- but not equal --- to the series part of the local solution in the original variable, e.g., the first term in $\LL^+$ can be written as $-\tau^{-2}\Liq_2(\tx Q\tx q,\tx q)$.

The shift operator $p$ acts on the dual variable as $p\tx Q = e^{2\pi i }\tx Q$, i.e., the terms in a power series in $\tx Q$ are invariant. The logarithm $\ln(\tx Q)$  in the dual variable is not invariant, but annihilated by the factor $(1-p)^2$ in the difference operator $\DL$ in eq.~\eqref{diffop} as well. I.e., each individual term at fixed $n$ in the sums $\LL^\pm$ is a solution to the difference equation and can be added to the perturbative $q$-periods  at order $H^2$ to obtain another solution.

Combining the terms in a local expansion near $Q=0$ or $w=0$ we obtain two new solutions to the original difference equation associated with the K-theory class $H^2$:
\begin{equation}\label{newsolH2}
\begin{aligned}
  \Pih_{0}(Q,\tx Q) & = \Liq_2(Q,q)+\LL^+(\tx Q,\tx q)+\LL^{(0)}(Q,q)\,,\\
  \Pih_{\infty}(w,\tx w) & = \Liq_2(wq^2,q)+\LL^-(\tx w,\tx q)\ ,
\end{aligned}
\end{equation}
with the additional parts adding a non-perturbative sector to the series \eqref{sol0} obtained from the perturbative 3d index. In the $S$-dual channel, the expansions around $Q=0$ and $w=0$ no longer cancel separately in the  analytic continuation as in eq.~\eqref{serexp1}, but 
\begin{equation}\label{acCFH2}
\Pih_{0}(Q)|_{H^2}+\Pih_{\infty}(w)|_{H^2}=0\ .
\end{equation}
A similar argument can be made for the analytic continuation of the coefficient of  $H^3$ in eq.~\eqref{IQexp}. 

\subsubsection{Cancellation of poles at fractional values of $\tau$}
Eq.\eqref{ac} implies trivialy the cancellation between poles in the local solutions and poles in the connection matrix $P$. Apart from $Q=q^k$, there is another interesting pole structure related to the special values of the complex structure $\tau$. The local expansions of the perturbative solution vector of the $q$-difference equation have, for general $X$, poles at $\tau\in\IQ$, i.e.,
\begin{equation}
  q^k=1\ ,\quad k\in\IZ\ ,
\end{equation}
In the above example these come from the series in eq.~\eqref{defi2}. 

It is easy to verify that the local solutions in the basis $\Pih$ that diagonalizes the connection matrix $P$ are regular at non-zero $\tau\in \IQ$ due to non-trivial cancellations, e.g. from poles in $\Liq_2(Q,q)$ and in $\LL^+(\tx Q,\tx q)$ in \eqref{newsolH2}. This follows immediately from the fact that the connection matrix is constant in this basis.\footnote{The case $\tau=0$ is special, since there are extra poles from the explicit factors of $\tau^{-1}$. These cancel only in the sum of the local expansions at $Q=0$ and $w=0$.}

Since the 3d GLSM studied in this section is supposed to be dual to the topological string, one expects a relation to a similar cancellation mechanism central to the proposal for the non-perturbative completion of the topological string in refs.~\cite{Hatsuda:2012dt,Hatsuda:2013gj,Hatsuda:2013oxa}. In the next section we explicitely verify that the substractions made to cancel the poles are identical for this special 3d GLSM, that is the subleading finite terms agree as well and \eqref{acCFH2} can be viewed as an analytic continuation in the non-perturbative topological string.

\subsubsection{Analytic structure on the $Q$-plane}
While the new solutions \eqref{newsolH2} are regular at $|q|=1$ and simplify the analytic continuation, they have more complictated analytic properties than the perturbative solutions.
The basic difference operators induced by 3d line operators act on the dual variables as
\begin{equation}\label{pshifts}
p:(Q,\tx Q)\to (Qq,\tx Qe^{2\pi i })\,, \quad \tx p:(Q,\tx Q)\to (Qe^{-2\pi i },\tx Q\tx q)\,,
\end{equation}
where $\tx p$ denotes a shift operator in the $S$-dual fugacity $\tx Q$. A loop around the origin in the $Q$-plane is linked to the insertion of a Wilson line operator in the dual sector, and similarly for a loop in the $\tx Q$ plane. These equations reflect the fact, that under the $S\in SL(2,\IZ)$ transformation of the 3d theory defined in ref.~\cite{Witten:2003ya}, a Wilson line in a global $U(1)$ symmetry is $S$-dual to a global vortex line \cite{Kapustin:2012iw}. 

Since  $[p,\tx p]=0$, the insertion of a dual Wilson line operator $\tx p$ gives another solution to the difference equation \eqref{diffop} in $p$. The result depends on the choice of $\ellf_Q$ in eq.~\eqref{pref}. For $\ellf_Q=\frac{\ln Q}{\ln q}$, the loop $Q\to e^{-2\pi i}$ generates a linear transformation on the perturbative solution vector \eqref{sol0}. On the other hand, for the factor \eqref{preftheta}, the terms at $H^0$ and $H^1$ are invariant, while the solutions at $H^2$ transform as
\begin{equation}\label{loopjumps}
  (\tx p^{k+1}-\tx p^k)\Pih_0=\displaystyle -\frac{1}{2\pi i \tau (1-\tx Q\tx q^k)}\ln(\tx Q e^{-2\pi i }\tx q^k)\ ,
\end{equation}
i.e., the non-perturbative solution is different on each sheet of the logarithmic $Q$ plane. In the limit $k\to \infty$ one obtains 
\begin{equation}
\lim_{k\to \infty}\Pih_0 = \LL^0+\Liq_2(Q,q)+\Delta(k)\ ,\quad \Delta(k)=-\frac{2\pi i }{(\ln q)^2}\ln(Qq^{-1}e^{-i\pi(k-1)})\ .
\end{equation}
The right hand side is a linear combination of the entries of the perturbative $q$-period vector~\eqref{sol0}, with a divergent coefficient for the terms in $\Delta(k)$.

\section{The Non-Perturbative Topological String}
We now verify that the non-perturbative corrections obtained from analytical continuation in the 3d GLSM match exactly the proposed non-perturbative completion of the topological string for the resolved conifold geometry \cite{Alim:2021ukq}. As a first check, one may relate the difference equation for the topological string to the difference equation~\eqref{diffop} for the 3d GLSM. For the resolved conifold the former reads~\cite{Alim:2020tpw,Alim:2021ukq}
\begin{equation}\label{diffeqtop}
  \Fgv(\lambda,t+\tfrac{\lambda}{2\pi})-2\Fgv(\lambda,t)+\Fgv(\lambda,t-\tfrac{\lambda}{2\pi})=\frac1{(2\pi)^{2}}\ln(1-e^{2\pi i t}) \ .
\end{equation}
Upon identifying $Q = e^{2\pi i t}$ and $q = e^{i\lambda}$ and defining with the difference operator $p$ in our conventions the potential $\Fgv(Q,q)$ as
\begin{equation}
     p \Fgv(Q,q) =(2\pi i)^2 \Fgv(\lambda,t) \ ,
\end{equation}
we arrive at the difference equation
\begin{equation}
  (1-p)^2 \Fgv(Q,q) =- \ln(1-Q) \ .
\end{equation}
The $H^2$ term of the $q$-period $\Pi_0$ in eq.~\eqref{sol0} corresponds to the logarithmic derivative
\begin{equation} \label{eq:FgvQq}
  {2\pi i} \,\partial_t\Fgv(\lambda,t-\tfrac{\lambda}{2\pi}) =  d_Q \Fgv(Q,q) \ ,
\end{equation}
which yields in the 2d limit the 4-cycle. Taking the logarithmic derivative of $\Fgv$ we get
\begin{equation}
  (1-p)^2 \left(d_Q\Fgv(Q,q)\right)=\frac{Q}{1-Q} \ .
\end{equation}
Adding the logarithmic term $\frac{\ellf_Q^2}2$ to $d_Q\Fgv(Q,q)$ changes the inhomogenous term of the difference equation to $(1-Q)^{-1}$. Therefore, $d_Q\Fgv(Q,q) +\frac{\ellf_Q^2}2$ is annihilated by the difference operator $(1-p)(1-Q)(1-p)^2$ and hence by $\DL=(1-p)^2(1-Q)(1-p)^2$, which is the difference operator of the 3d GLSM in eq.~\eqref{diffop} for the Chern--Simons levels $(a,b)=(0,0)$. The additional factor $(1-p)$ in $\DL$ allows for the extra solution for the coefficient of $H^3$ in the $q$-period~\eqref{sol0}.

The non-perturbative topological string partition function $\Fgv^\text{np}(\lambda,t)$ is related to a variant of the triple sin function appearing in refs.~\cite{Lockhart:2012vp,Bridgeland:2016nqw} according to ref.~\cite{Alim:2021ukq}
\begin{equation}
   \Fgv^\text{np}(\lambda,t)=\ln G_3(t|\tfrac \lambda {2\pi},1) \ .
\end{equation}
The right hand side has the contour integral representation \cite{MR2101221,Bridgeland:2016nqw,Alim:2021ukq}
\begin{equation}\label{lnG3}
  \ln G_3(t|\omega_1,\omega_2) = - \int_{\mathbb{R} + i0_+} \!\!
  \frac {ds}s \frac {e^{(t+\omega_1)s}}{(e^{\omega_1s}-1)^2(e^{\omega_2s}-1)}\ ,
\end{equation}
with $0<\operatorname{Re}\omega_1$, $0<\operatorname{Re} \omega_2 $, $-\operatorname{Re}(\omega_1)< \operatorname{Re}(t) < \operatorname{Re}(\omega_1 + \omega_2)$, and the contour is taken along the real axis encircling the pole at $s=0$ in the positive imaginary plane. Applying the logarithmic derivative $d_Q$ with eq.~\eqref{eq:FgvQq} yields the integral representation
\begin{equation}
    d_Q \Fgv(Q,q) = - \frac{1}{2\pi i} \int_{\mathbb{R} + i0_+} \!\! ds \frac{{Q}^\frac{s}{2\pi i} }{(q^\frac{s}{2\pi i} -1)^2 (e^s-1)} \ ,
\end{equation}
for $\arg q > 0$. Evaluating this integral for $|Q|<1$ by closing the contour with an arc at infinity in the upper Siegel half plane, the integral is readily computed by summing up the residues of the simple poles at $s= 2\pi i k$ and the double poles at $s=\frac{(2\pi i)^2}{\ln q} k$ for positive integers $k$. The former poles yield the function $\Liq_2(Q,q)$ defined in eq.~\eqref{defi2}, whereas the latter poles realize the function $\LL^+(\tx Q,\tx q)$ in the $S$-dual variables $\tx Q$ and $\tx q$, such that we obtain altogether
\begin{equation}
   d_Q \Fgv(Q,q) = \Liq_2(Q,q) + \LL^+(\tx Q,\tx q) \ .
\end{equation}   
which --- up to the logarithmic terms $L_2^{(0)}(Q,q)$ --- coincides with the non-perturbative period of the K-theory class $H^2$ given in eq.~\eqref{newsolH2}.

\section{Outlook}
In this note the double periodicity and meromorphicity of the entries of the Birkhoff connection matrix is used to compute non-perturbative corrections to the index of the 3d GLSM with geometric Higgs phase $X$ by analytic continuation. For the special 3d GLSM dual to the topological string on the conifold, the method reproduces the existing proposals on a non-perturbative definition, and the computation is a check on the duality. The new aspect is the relation between the expressions given for the topological string in the literature, and the elliptic functions in the connection matrix.

The method is applicable more generally to compute non-perturbative corrections to the 3d GLSM for compact Calabi--Yau target space $X$, e.g., the theories described in Section~\ref{sec:GLSM}. In this case one cannot expect the result to correspond directly to a non-perturbative topological string partition function for the Calabi--Yau space $X$ on several grounds,\footnote{For recent progress on studying the non-perturbative topological string on compact Calabi--Yau threefolds, see refs.~\cite{Gu:2023mgf}.} notably the more complicated RG flow. Since the information in the perturbative 3d index for the Calabi--Yau target space $X$ can be rephrased in terms of the Gopakumar--Vafa invariants for $X$ \cite{Jockers:2019wjh}, there should still be an interesting connection between the non-perturbative completions of the two sides. 

It would be interesting to formulate from the proposed non-perturbative completion of the analyzed index of the 3d GLSM, a definition of a more fundamental theory. For the special 3d GLSM studied in this note, this more fundamental theory should be dual to the topological string on the conifold. I hope to come back to this important issue in the future elsewhere.

\bigskip
\subsection*{Acknowledgments} 
Foremost, I am deeply thankful to Peter Mayr for initially collaborating with me on this work. Without his insight this work would not have been possible.
I would like to thank Murad Alim, Mykola Dedushenko, Duco van Straten and Stavros Garoufalidis for discussions and correspondences. 
This work is supported by the Cluster of Excellence Precision Physics, Fundamental Interactions, and Structure of Matter (PRISMA+, EXC 2118/1) within the German Excellence Strategy (Project-ID 390831469).

\bigskip
\bibliographystyle{utphys}
\bibliography{JoNonPert}
\end{document}